\documentclass[]{aa}
\usepackage{amsmath}
\usepackage[varg]{txfonts}
\usepackage{overpic}

\usepackage{natbib}
\bibpunct{(}{)}{;}{a}{}{,} 
\usepackage{graphicx}
\usepackage{placeins}

\usepackage{color}
\usepackage{bm}
\usepackage{esdiff}
\usepackage{amssymb}
\usepackage{amsbsy}
\usepackage{fontenc}
\usepackage{overpic}
\usepackage{multirow}

\authorrunning{S. Th\"olken et al.}

\title{
Discovery of large scale shock fronts correlated with the radio halo and radio relic in the A2163 galaxy cluster
}

\titlerunning{Discovery of large scale shock fronts in the A2163 galaxy cluster}

\author{Sophia Th\"olken\inst{\ref{inst1}}\
\and Thomas H. Reiprich\inst{\ref{inst1}}\
\and Martin W. Sommer\inst{\ref{inst1}}\
\and Naomi Ota\inst{\ref{inst2}}\
}

\institute{Argelander-Institut f\"ur Astronomie, Universit\"at Bonn, Auf dem H\"ugel 71, 53121 Bonn, Germany\label{inst1}\\
\email{thoelken@astro.uni-bonn.de}  \and
Department of Physics, Nara Women’s University, Kitauoyanishi-machi, Nara, Nara 630-8506, Japan\label{inst2}}

\date{Received date /
Accepted date }

\abstract {Galaxy clusters form at the intersections of the filamentary large scale structure in merging events and by the accretion of matter along these filaments. Imprints of these formation processes should be visible in the intracluster medium and can arise in shock fronts, which are detectable via discontinuities in e.g. the gas temperature and density profiles. However, relatively few observational examples of prominent shocks are detected in X-rays so far.} 
{In this study, we investigate the X-ray properties of the intracluster gas and the radio morphology of the extraordinary cluster A2163. This cluster shows an irregular morphology in various wavelengths and has one of the most luminous and extended known radio halos. Additionally, it is one of the hottest clusters known. We aim for measuring the temperature and density profiles in two azimuthal directions to search for the presence of shock fronts.} 
{We perform a spectral analysis of data from two Suzaku observations, one in the north-east (NE) and one in the south-west (SW) direction of A2163, and use archival XMM-Newton data to remove point sources in the field of view. We deproject the temperature and density profiles and account for the Suzaku point-spread-function. From the detected discontinuities in the density and temperature profiles, we estimate the Mach numbers and velocities of the shock fronts. To compare our findings in the X-ray regime with the radio emission, we obtain radio images of the cluster from an archival VLA observation at $20$\,cm.} 
{We identify three shock fronts in A2163 in our spectral X-ray study. A clear shock front lies in the NE direction at a distance of $1.4$\,Mpc from the center, with a Mach number of $M=1.7_{-0.2}^{+0.3}$, estimated from the temperature discontinuity. This shock coincides with the position of a known radio relic. We identify two additional shocks in the SW direction, one with  $M=1.5^{-0.3}_{+0.5}$ at a distance of $0.7$\,Mpc, which is likely related to a cool core remnant, and a strong shock with $M=3.2_{-0.7}^{+0.6}$ at a distance of $1.3$\,Mpc, which also closely matches the radio contours. The complex structure of A2163 as well as the different Mach numbers and shock velocities suggest a merging scenario with two unequal merging constituents, where two shock fronts emerged in an early stage of the merger and traveled outwards while an additional shock front developed in front of the merging cluster cores.}{}

\keywords{galaxies: clusters: general - galaxies: clusters: individual: A2163- X-rays: galaxies: clusters - radio continuum: general}

\def\Vhrulefill{\leavevmode\leaders\hrule height 0.7ex depth \dimexpr0.4pt-0.7ex\hfill\kern0pt}

\begin{document}

\maketitle

\section{Introduction}

At the intersections of the filamentary large scale structure, the largest distinct building blocks, the galaxy clusters, are located (e.g.\ \citealp{2001Natur.409...39B}, \citealp{2005Natur.435..629S}, \citealp{2014MNRAS.444.1518V}). Merging events of galaxy clusters and groups are the most energetic processes in the universe and a significant fraction of the energy is dissipated into shock heating of the intracluster medium (ICM) (e.g. \citealp{2018ApJ...857...26H}, \citealp{2015ApJ...812...49H}, \citealp{2007MNRAS.376..497M}). These shock fronts are visible in X-ray observations of the ICM as discontinuities in the temperature, density and surface brightness (SB) profiles.  A prominent example for a distinct merger shock is the Bullet cluster (\citealp{2002ApJ...567L..27M}) with a large Mach number of $M{\sim}3$ (\citealp{2007MNRAS.380..911S}). Often, cold fronts are observed in mergers (see \citealp{2010A&A...516A..32G} and references therein), caused by the cool core remnants of the merging constituents, while the actual shock front is located in front of the cold front, as it is also the case for the Bullet cluster.

Until now, relatively few detailed X-ray studies of shocks in the ICM have been performed. For example, \citet{2009A&A...495..721S} studied the merging cluster Hydra A using deep XMM-Newton observations and found a large scale shock, visible in the surface brightness profile and pressure map, with a Mach number of about $1.3$. This is a typical value of X-ray detected shocks which usually have $M\lesssim3$ (\citealp{2007PhR...443....1M}).  

{A recent study by \citet{2018MNRAS.476.5591B} searched for discontinuities in a sample of 15 clusters with Chandra and found six shock and eight cold fronts and several discontinuities with uncertain origin. All the detected shocks have Mach numbers smaller than 2, estimated from  the temperature and density profiles.}

Another example, studied by \citet{2013ApJ...764...82B} with XMM-Newton, is the cluster A521. This cluster exhibits two cold fronts which separate the merging constituents and two shock fronts propagating outwards in east and south-west direction. The shock heated regions and one of the shock fronts show spatial correlations with a radio halo and a radio relic, respectively. 
Radio halos are typically associated with disturbed/merging clusters, suggesting that the acceleration of electrons, responsible for the radio emission, is produced by turbulence in the intra-cluster medium following a merger (e.g. \citealp{2017ApJ...843L..29E}, \citealp{2011MNRAS.412..817B}, \citealp{2007MNRAS.378..245B}).
Other examples of shocks have e.g. been observed by \citet{2011ApJ...728...82M}, \citet{2012PASJ...64...67A}, \citet{2017A&A...600A.100A} and \citet{2017PASJ...69...39H},  for the clusters A754, A3376, A2255 and A2744, respectively. In all cases, radio emission is observed and the X-ray detected shocks seem to be related to radio relics. For A3376 and A2744, the authors report high Mach numbers of $M{\sim}3$ and $M=3.7\pm0.4$, respectively, estimated from the gas temperature jump. 

A review about X-ray detected shocks in the ICM can be found in \citet{2007PhR...443....1M}.

 \citet{2014MNRAS.440.3416O} detected multiple temperature and density discontinuities in the merging cluster CIZA J2242.8+5301 using Chandra and Suzaku data. This cluster hosts a double radio relic system and the radio derived Mach number is almost a factor of two larger then the Mach number derived from the Suzaku X-ray temperature profile. This discrepancy between radio- and X-ray-derived Mach numbers is a known conundrum. \citet{2015ApJ...812...49H} analyzed cosmological hydrodynamic simulations, adopting the diffusive shock acceleration (DSA) model, and extracted radio and X-ray shock properties in the 2D observer plane. They found, that projection effects along the line of sight have a significant impact and radio observations tend to detect stronger shocks while X-ray derived Mach numbers are typically lower.   \newline


In this work, we study the disturbed cluster A2163 ($z = 0.203$, \citealp{1999ApJS..125...35S}), which is one of the hottest known clusters with a gas temperature of $T=14.6_{-0.8}^{+0.9}$\,keV (\citealp{1995A&A...293..337E}) and also shows prominent radio emission (\citealp{2001A&A...373..106F}). We analyze two Suzaku observations in the north-east (NE) and south-west (SW) direction and investigate the ICM properties with respect to the presence of possible shock fronts. 

Previous detailed studies of A2163 in the radio and optical band revealed a complex merging situation with several substructures and likely two merging constituents. \citet{2011ApJ...741..116O} performed a weak lensing analysis of A2163 and found a bimodel mass distribution. The mass distribution coincides with the galaxy density distribution but seems spatially offset from the brightest X-ray emitting regions, which the authors interpret in terms of ram pressure stripping of the gas in the merging process.  \citet{2001A&A...373..106F} identified one of the largest known radio halos in A2163, which shows spatial correlations to the X-ray surface brightness. Their radio analysis showed the presence of a radio structure in the NE direction which was classified as a radio relic. Later, \citet{2004A&A...423..111F} investigated the spectral index map of A2163, which supported the interpretation that the previously identified radio feature is indeed a radio relic. \citet{2008A&A...481..593M} compared the findings in the radio regime with the galaxy density and found an even more pronounced correlation and several subclumps with clear galaxy overdensities. 

An XMM-Newton and Chandra X-ray study of A2163 was performed by \citet{2011A&A...527A..21B}. Their findings, obtained from spectral imaging, suggest that the ICM is in a highly disturbed status with a hotter region in the NE direction and a cool core remnant, located SW of the cluster center. In their interpretation, the cool core was likely separated from its host halo. A galaxy overdensity close by (\citealp{2008A&A...481..593M}) lends support to this interpretation. 
The cool core is likely moving westwards and also exhibits a cold front as found by \citet{2011A&A...527A..21B} using Chandra data. The authors also identified a prominent structure north of the cluster center, which we also see in our Suzaku observation (see Fig. \ref{fig:pointings}), however, they state that no indications for an interaction with the main cluster are found.

\citet{2014A&A...562A..60O} looked for hard X-ray emission in A2163 using Suzaku HXD data. 
They find significant emission in the $12-60$\,keV band and a multi-temperature structure with a high temperature component in the NE direction, but no significant non-thermal X-ray emission. 

A detailed comparison of our work to these findings and possible interpretations are given in Sec.~\ref{sec:discussion}. 

The extensive multi-wavelength studies of A2163 brought up interesting results. However, despite its disturbed appearance, no shock fronts have been detected so far, although their presence seems likely given the clusters substructured morphology. With our Suzaku analysis, we aim to identify these shocks and we are able to robustly probe the gas properties out to larger radii than what is possible with XMM-Newton or Chandra, due to Suzaku's low and stable instrumental background.

Throughout the analysis we assume $\Omega_\Lambda = 0.7$, $\Omega_{\rm m} = 0.3$, $H_0 = 70$\,km/s/Mpc which corresponds to a scale of 3.3\,kpc/$''$ at the cluster redshift of $z=0.203$. All uncertainties are quoted at the 68\% level.

\section{Observations and data reduction}\label{sec:obs}

\subsection{Radio data}
The radio halo in A2163 was first studied by \citet{2001A&A...373..106F},
using the same Very Large Array (VLA) data that we use here.

A2163 was observed at $20$\,cm wavelength with the VLA
C and D configurations under project numbers AF328 in 1998 and 1999,
using central frequencies of 1.36\,GHz and 1.46\,GHz with 25\,MHz bandwidth
per frequency in the C configuration and 50\,MHz bandwidth per
frequency in the D configuration. The on-source integration time was
3.9\,h (C configuration) and 3.4\,h (D configuration).

\subsection{X-ray data}

In this work, we analyze two Suzaku XIS observations of A2163 in NE and SW direction of 39\,ks and 109\,ks cleaned exposure time, respectively. Details of the observations can be found in Tab. \ref{tab:observations}. Additionally, we use archival XMM-Newton data (PI: H. Bourdin, Obs.-ID: 069450010) to detect point sources in the field of view.
We follow the standard Suzaku data reduction procedure, which includes the following steps: Assign coordinates, time and pixel quality flags to each event, apply gain- and charge-transfer-inefficiency correction and identify anomalous pixels. For the geomagnetic cut off rigidity we use a limit of COR2$>$6 and events falling in the second trailing rows of the charge injection rows are discarded.

\begin{table*}
\caption[Details on the analyzed observations of A2163]{Details of the Suzaku observations of A2163. The exposure time is the cleaned exposure time after data reduction.}             
\centering      
\small
\begin{tabular}{c c c c c c}     
\hline \hline                      
 & Date & Pointing (R.A., Dec.) & Exp. Time & Obs-ID & PI\\ \hline
South-west (SW) & 2008 Aug & (243.8071, -6.2200) & 109.2\,ks & 803071010 & T. H. Reiprich \\ 
North-east (NE) & 2009 Feb & (244.0216, -6.0449) &~38.6\,ks  & 803022010 & N. Ota \\ 
\hline \hline                                             
\end{tabular}
\label{tab:observations}
\end{table*}

\begin{figure}[htbp]
\centering
\includegraphics[width=0.49\textwidth]{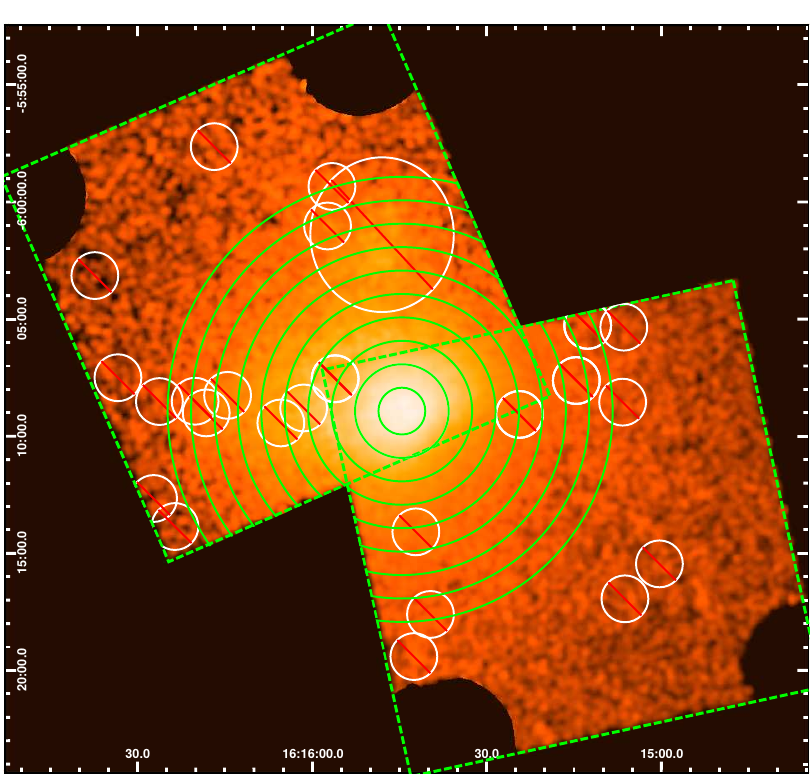}
\caption[Suzaku pointings and extraction regions for A2163]{Suzaku pointings (green dashed boxes), extraction regions for the Suzaku analysis (green annuli) and removed (point source) regions (white circles). Point sources are removed above a flux limit of $3\times10^{-14}$\,erg/s/cm$^2$. As in  \citet{2011A&A...527A..21B}, the structure in the north is removed because it seems unrelated to the main cluster.}
\label{fig:pointings}
\end{figure}

Suzaku observations can be contaminated by solar flares, which cause solar-wind-charge-exchange emission. We perform a flare filtering by applying a three-sigma clipping to the light curves and removing the corresponding time intervals. Point sources are detected using the archival XMM-Newton observation, which we also clean for flares and which covers the analyzed area in both Suzaku pointings. The task {\it emldetect} is used to detect point sources in the field of view in two energy bands from $0.5-4.5$\,keV and $4.5-12.0$\,keV. We apply the energy conversion factors from \citet{2016A&A...590A...1R}. Not all of the detected sources can be removed from the Suzaku observation as this would remove a significant amount of area due to Suzaku's larger point-spread-function (PSF). Thus, we choose a flux limit of $3\times10^{-14}$\,erg/s. In the NE direction, we identify the same substructure as found by \citet{2011A&A...527A..21B}, which we exclude from the analysis. All removed areas are shown in Fig. \ref{fig:pointings}.

\section{Analysis}\label{sec:analysis}

\subsection{Radio analysis}
The interferometric radio data are calibrated using the Common
Astronomy Software Applications (\citealp{2007ASPC..376..127M})
package. Primary flux calibration is performed against the known
calibrator 3C286, adopting the flux scale of
\citet{2013ApJS..204...19P}. Phase calibration is carried out against
bright nearby compact sources (that were found to be observed
approximately every thirty minutes) for which we in turn determine
gain solutions from the primary calibrator.

We perform several iterations of self-calibration, for each antenna
configuration separately, to improve upon the phase solutions. Upon
combining the data from the two configurations, we perform one
iteration of amplitude self-calibration, with solutions averaged over
one hour, followed by one final iteration of phase
self-calibration. Residual phase errors are found to be less than a
few degrees across all antennas.

To image the radio halo, we first make a model image of the compact
sources in the field, using only baselines longer than $1.8$\,k$\lambda$
(corresponding to around $380$\,m at $20$\,cm wavelength) and
multi-frequency deconvolution \citep{2011A&A...532A..71R} with two
Taylor terms to model spectral slopes in the continuum. De-gridding
the compact source model to \textit{uv} space, we subtract the model
from the visibilities and proceed to image the radio halo using the
short interferometer spacings, which are sensitive to the extended
emission. To image the extended emission, we use Briggs weighting
with ${\rm robust}=1$ and a \textit{uv} taper at  $45'$. Due to the large size
of the radio halo, we use multi-scale CLEANing to deconvolve the
image. Our final image (Fig. \ref{fig:radio_img}) has a reconstructed circular Gaussian beam of
$50''\times 50''$ at full-width half-maximum. Structure is recovered to scales
of approximately $12'$, which is close to the limit of what can be
imaged with the VLA.

\subsection{X-ray analysis}
\subsubsection{Background}
The Suzaku non-X-ray background (NXB) is caused by highly energetic particles hitting the detector and producing continuum and fluorescent line emission. This background is estimated from night earth observations in a time interval of $\pm 150$ days around the observation date and subtracted from the source spectra. For the creation of the NXB spectra, we followed the improved treatment of the flickering pixels\footnote{see {heasarc.gsfc.nasa.gov/docs/suzaku/analysis/xisnxbnew.html}}.

The X-ray background is mainly composed of three components: a local component from the local hot bubble and solar wind charge exchange (this component is called LHB in the following), a milky way halo component and the superposition of distant AGNs causing a diffuse background (called CXB in the following). We model these components using an {\it apec} and {\it absorbed apec} model for the LHB and the halo component, respectively, and an {\it absorbed power law} for the CXB component. The LHB temperature is fixed to 0.1\,keV and the CXB component has a photon index of 1.41 (\citealp{2004A&A...419..837D}). The redshift of the {\it apec} models is fixed to 0 and the abundance is set to 1. The temperature of the halo component is fixed to 0.28\,keV (e.g., \citealp{2010PASJ...62..371H}, \citealp{2011PASJ...63S1019A}). In addition to the Suzaku data, we use ROSAT All-Sky Survey data to constrain the background model parameters in a region far off the center where no cluster emission is expected\footnote{Spectra were extracted with the HEASARC X-ray background tool {heasarc.gsfc.nasa.gov/cgi-bin/Tools/xraybg/xraybg.pl}}. 

We create ancillary response files for the background spectra using $xissimarfgen$ and assuming a uniform distribution.

\subsubsection{PSF correction}\label{sec:psfcorr}
The limited spatial resolution of Suzaku requires to correct for the effect of the PSF, which is  $1.8'$ half-power-diameter. Therefore, we simulate the mixing of photons between the different annuli for each of the observations separately with {\it xissim}. As in \citet{2016A&A...592A..37T}, we use an iterative approach for the input image of the simulation. In a first step, we obtain background-subtracted SB profiles from Suzaku images in the NE and SW direction, which we fit by single-beta models and use these models as inputs to the simulation. 
%
From this simulation, we obtain PSF-correction factors which are introduced in the fitting procedure. 
From this spectral fit, PSF-corrected SB profiles are obtained, which are then again used as inputs to the simulation and the analysis is repeated.
For details about this procedure see \citet{2016A&A...592A..37T}. To reduce the correlations in the fit due to the PSF correction, we account for PSF mixing factors $\geq 10\%$ as we do not expect strong influence from annuli with lower mixing factors.  However, significant correlations between neighboring annuli are still present, which are found to result in temperature ``oscillations" in the fit. We thus regularize the temperature and abundance values by allowing them to vary within $\pm 2\sigma$ of the values of a PSF-uncorrected fit. However, the normalization is not constrained and we 
discuss the effect of the regularization on our results in Sec. \ref{sec:shock}. 

\subsubsection{Fitting procedure}
We analyze the observations in the NE and SW direction separately and extract spectra in 10 and 9 radial bins, respectively, each with $1'$ width. We examine the signal-to-background ratio ($N_{\rm source}/N_{\rm bkg}$) in each annulus and limit the number of regions such, that the ratio is ${\sim} 1$ in the outermost annulus. The extraction regions are shown in Fig. \ref{fig:pointings}.

The source emission is modeled by an {\it absorbed apec} model with a fixed redshift of $z = 0.203$ and the column density taken from \citet{2013MNRAS.431..394W}, which includes molecular hydrogen. 
{Since A2163 lies behind a relatively dense hydrogen cloud, we also tested the impact on the temperature profile when leaving the column density in each spectral annulus as a free fit parameter, but found only marginal differences in each bin.}

The abundance table from \citet{2009ARA&A..47..481A} is used throughout this analysis. The metal abundance is estimated in four radial bins, i.e. this parameter is linked across several annuli according to Tab. \ref{tab:fitresults_NE} and Tab. \ref{tab:fitresults_SW}, for the NE and SW direction, respectively.


Ancillary response files for both observations are created using the same iterative approach for the beta-model profiles as used for the PSF simulation in Sec. \ref{sec:psfcorr}.
All spectra are grouped to at least 25 counts per bin and analyzed in the energy range $0.7-10$\,keV, except for the outermost three (two) annuli in the NE (SW) direction, for which the fitting range is $0.7-7$\,keV to achieve a sufficient signal-to-background ratio.

\subsubsection{Deprojection}\label{sec:deprojection}
%
%
%
%

To deproject our results, we follow the method described in \citet{2016A&A...592A..37T}, based on \citet{2002MNRAS.331..635E}, and assume piecewise functions to model the electron density and temperature profiles with the discontinuities caused by the shock fronts. The temperature and density profiles are deprojected simultaneously and correlations between the data points due to the PSF correction are taken into account using the covariance matrix from the spectral fit.
The irregular morphology of the cluster makes it necessary to treat the NE and SW direction separately as described in the following. 

The temperature model in the NE direction is described by a simple step function
\begin{equation}\label{eq:deprojmodelT_NE}
 T^{\rm NE}(R) = \begin{cases} T_0^{\rm NE},\quad R\leq R_{\rm j} \\
         T_1^{\rm NE},\quad R>R_{\rm j}
        \end{cases},
\end{equation}
with $R_j$ being the position of the discontinuity. The electron density in NE direction is, correspondingly, modeled by a piecewise power law
\begin{equation}
\begin{split}
 n^{\rm NE}_{\rm e}(R) = \begin{cases} n_1 R^{-\eta_1},&\quad R\leq R_j \\
 n_2 R^{-\eta_2}, &\quad R>R_j
\end{cases}\end{split}\quad ,
\end{equation}
with normalizations $n_1$ and $n_2$, and slope parameters $\eta_1$ and $\eta_2$. 
This simplified density model is not able to reproduce the central part of the cluster for which usually a (double) beta-model is needed. Thus, we exclude the central bin in the deprojection.

In the SW direction, a more complex modeling is necessary. The SW temperature profile is discussed in detail in Sec. \ref{sec:results} and we anticipate the results here to argue for our used deprojection models. In a first test, we applied the same temperature and density models for the deprojection as used in the NE direction. However, with these models, we find a bad description of our measured quantities, 
($\chi^2$/d.o.f$=3.6$)
which seem to require more advanced deprojection models. 
Nevertheless, this first test with the simple single-jump models indicates, that a shock position around ${\sim}3.5'$ is preferred instead of the clearly visible jump in the SW temperature profile at around $6'$ (see Sec. \ref{sec:results}, Fig. \ref{fig:temp_profiles}). 
For this reason, we test models which allow for two discontinuities at the positions $R_{\rm j,1} < R_{\rm j,2}$. In the temperature profile, the first and second position are smoothly connected via a Sigmoid function. These models yield a clearly better description of our measured values in the SW direction, regarding the reduced Chi-squared ($\chi^2$/d.o.f$=0.8$) and are therefore used in the following. 
Using the F-test, the single-jump model is rejected with a false-rejection probability of less than 10\%. Thus, the final SW temperature model is given by
\begin{equation}\label{eq:temp_profile_sw}
 T^{\rm SW}(R) = \begin{cases} T_0^{\rm SW},\quad R\leq R_{\rm j,1} \\
	S(R),\quad R_{\rm j,1}<R\leq R_{\rm j,2} \\
         T_2^{\rm SW},\quad R>R_{\rm j,2}
        \end{cases},
\end{equation}
with the Sigmoid function
\begin{equation}\label{eq:sigmoid}
 S(R) =  \frac{(T_{1,b}-T_{1,a})}{1+\exp{\left(-m[R - 0.5(R_{\rm j,2} - R_{\rm j,1})]\right)}} + T_{1,a},
\end{equation}
which models the smooth transition between the two temperatures at the jump positions $T_{1,a} < T_{1,b}$ with slope $m$. The turning point is chosen to lie at half distance between the two jumps.

The density is, correspondingly, described by
\begin{equation}\label{eq:deprojmodeln_SW}
 n_{\rm e}^{\rm SW}(R) = \begin{cases} n_0^{\rm SW}R^{-\alpha},\quad R\leq R_{\rm j,1} \\
	n_1^{\rm SW}R^{-\beta},\quad R_{\rm j,1}<R\leq R_{\rm j,2}\\
	n_2^{\rm SW}R^{-\gamma},\quad R>R_{\rm j,2}
        \end{cases}.
\end{equation}


%
%

%


\section{Results}\label{sec:results}

\subsection{Radio emission}
\begin{figure*}[htbp]
\centering
\includegraphics[width=0.99\textwidth]{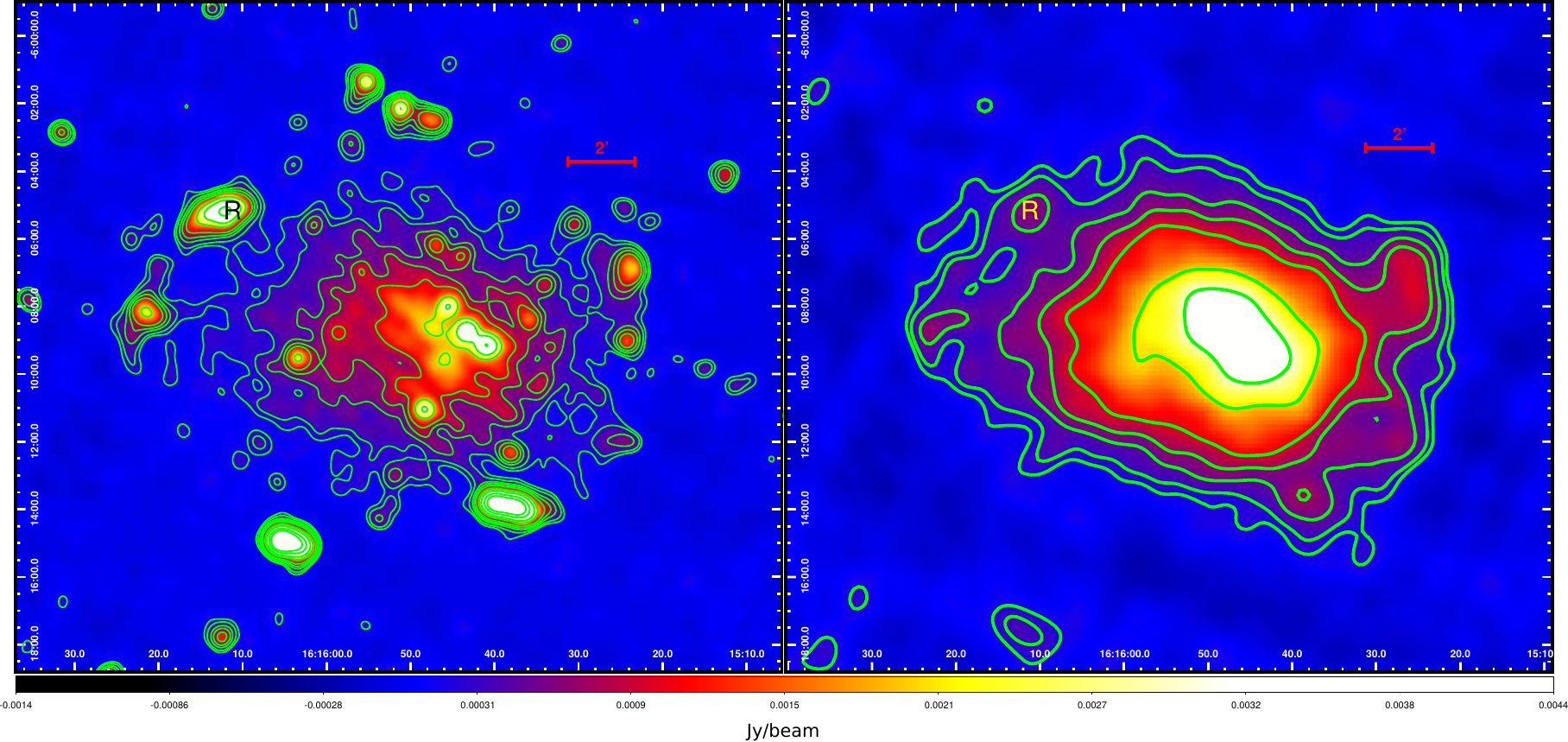}
\caption[]{Radio images of A2163 at 20\,cm. The contour levels are 0.2, 0.3, 0.5, 0.7, 1, 2, 3, 5, 7\,mJy/beam. The letter ''R`` indicates the position of the radio relic, as identified by \citet{2001A&A...373..106F}. {\it Left:} Radio map with a circular Gaussian beam size of $30''$ FWHM. The noise level is 0.05\,mJy/beam. {\it Right:} Image of the extended emission after the removal of compact sources with a circular reconstructed beam size of $50''$ FWHM. The noise level is 0.1\,mJy/beam.}
\label{fig:radio_img}
\end{figure*}

Fig. \ref{fig:radio_img} shows the radio images obtained from our VLA analysis. As identified previously by \citet{2001A&A...373..106F}, A2163 exhibits several compact sources, a large extended radio halo and a radio relic in the NE. The latter is supported by measurements of the spectral index map between 0.3 and 1.4\,GHz by \citet{2004A&A...423..111F}. The radio emission is slightly elongated in E-W direction which supports the interpretation by \citet{2011A&A...527A..21B} and \citet{2008A&A...481..593M} that the merging is proceeding along this direction. A detailed comparison of the radio and X-ray results is presented in Sec. \ref{sec:discussion}. 

\subsection{X-ray temperature and emission measure profiles}

\begin{figure}[htbp]
\centering
\includegraphics[width=0.49\textwidth]{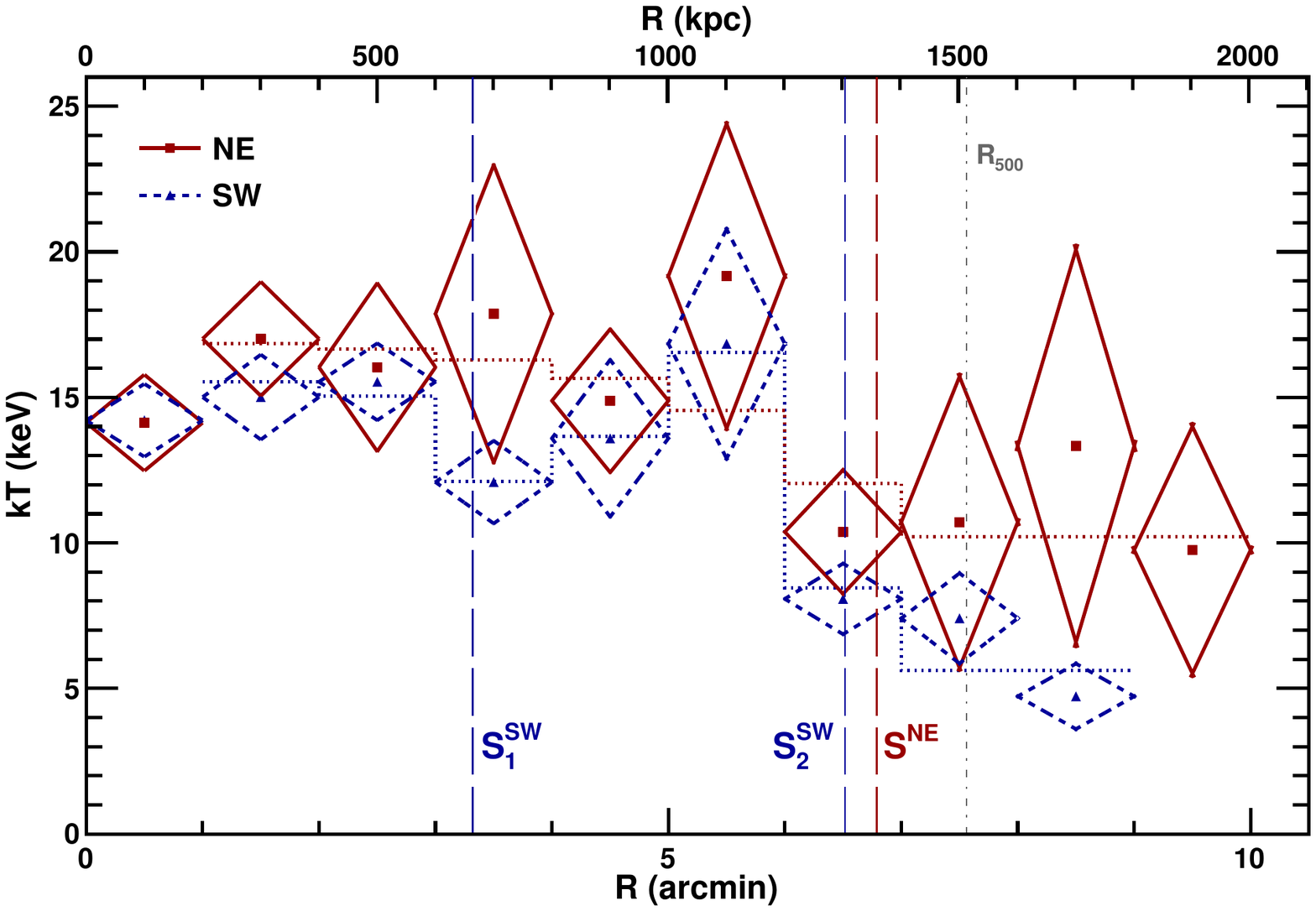}
\caption[Temperature profiles of A2163]{Projected, PSF-corrected Suzaku temperature profiles of A2163 obtained in NE (solid red diamonds) and SW (dashed blue diamonds) direction. The vertical dashed lines show the jump positions of the detected shocks as obtained from our deprojection procedure while dotted lines correspond to the projected best-fit temperature models (see Sec.~\ref{sec:deprojection} for details about the deprojection). The gray vertical dot-dashed line indicates $R_{500}$, as obtained by \citet{2008A&A...487...55R} from a weak-lensing analysis.}
\label{fig:temp_profiles}
\end{figure}

Fig. \ref{fig:temp_profiles} shows the measured, PSF-corrected temperature profiles of A2163 in the NE and SW direction as well as the projected model temperature values, which are obtained by projecting the best-fit model of the deprojection procedure along the line of sight. The projected models are able to well reproduce the observed profiles. The obtained temperatures are given in Tab. \ref{tab:fitresults_NE} and \ref{tab:fitresults_SW}, for the NE and SW direction, respectively. As expected, the profiles agree in the inner cluster region, where the two Suzaku pointings overlap, but show clear deviations farther out. In general, the temperature in SW direction is lower compared to the NE direction, which has a relatively flat temperature profile. Most notably, both profiles exhibit a clear temperature drop around $6.5'$, {which was also seen in an early study by \citet{2001ApJ...563...95M} using a short Chandra observation, albeit with large uncertainties due to their short exposure.}


\begin{figure}[htbp]
\centering
\includegraphics[width=0.49\textwidth]{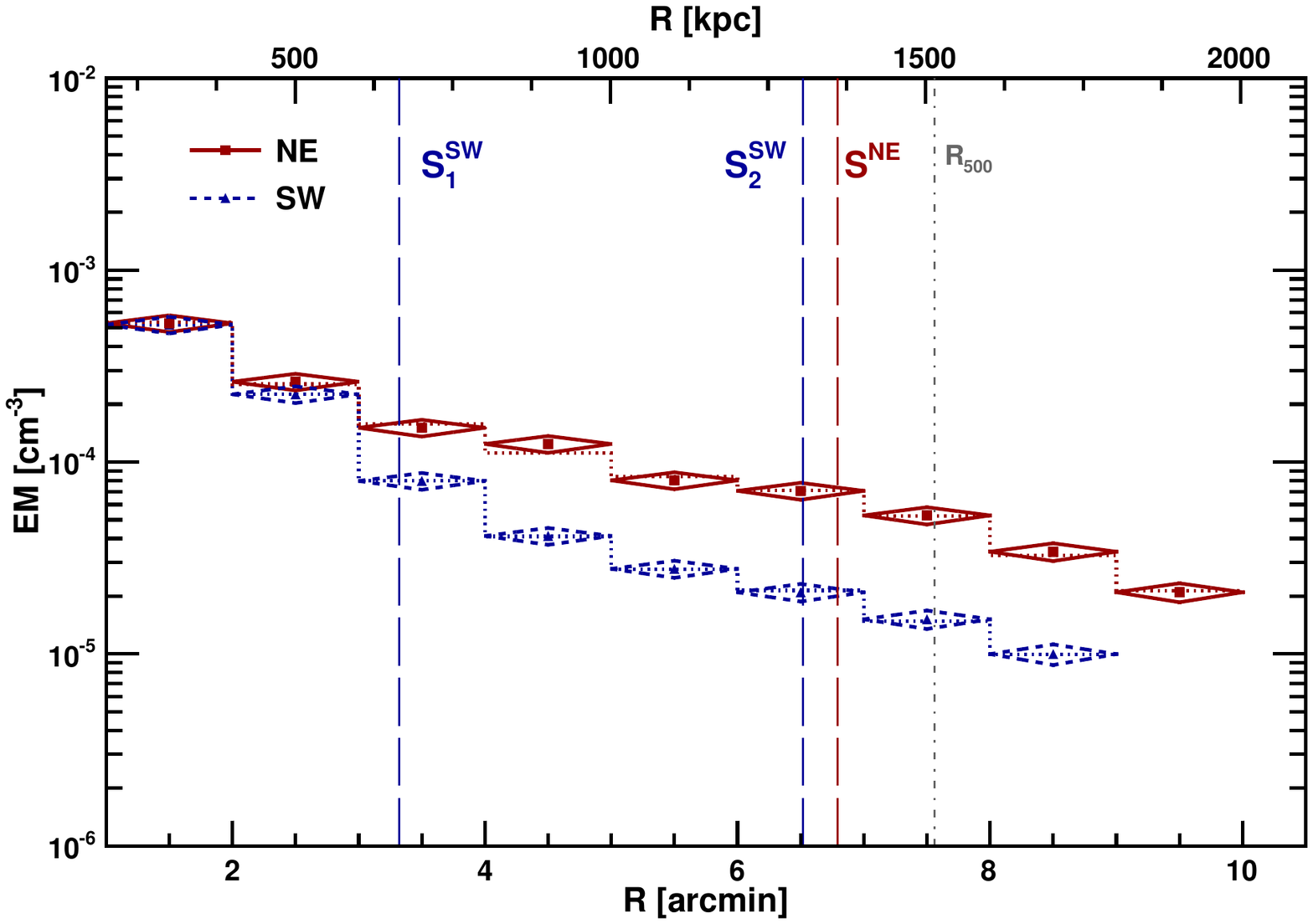}
\caption[Emission measure profile of A2163]{Projected, PSF-corrected emission measure profile of A2163 obtained from the spectral fit for the NE (solid red diamonds) and SW (dashed blue diamonds) direction. Dashed and dotted lines have the same meaning as in Fig.~\ref{fig:temp_profiles}. The innermost annulus is not shown as it is not considered in the deprojection procedure (see text). 
The profiles reproduced from the deprojection models (dotted lines) well match the measured profiles.
}
\label{fig:em_profile}
\end{figure}

In the SW direction, we measure an increase in temperature between the fourth and sixth radial bin, followed by a sudden drop from ${\sim}17$\,keV to ${\sim}8$\,keV, which is a clear indication of a shock front. With our deprojection algorithm, we identify a second discontinuity located at ${\sim}3.3'$. By eye, an indication for this discontinuity in the measured temperature profile can be seen in the fourth annulus, where the temperature slightly drops. Also the emission measure profile, shown in Fig. \ref{fig:em_profile}, shows a stronger drop around that radius than the NE direction and is overall clearly flatter than the SW profile. 
{No edges are detected in the emission measure profile around the positions of the outer shocks. We therefore also checked the SB profile of the XMM-Newton observation but found no associated features, which is somehow surprising given the clear temperature discontinuities. This has also been seen before by e.g. \citet{2017A&A...600A.100A} who also could not identify a SB edge for their detected shock front despite a clear temperature jump across a radio relic. Interestingly, the SB profile of A2163 of an early ROSAT study (\citealp{1995A&A...293..337E}) shows a slight hint for a feature around the outer shock positions, which is, however, not seen in our analysis and deeper XMM-Newton or Chandra data might be required to further investigate this point. In the future, eROSITA might also help resolving this issue with its very good soft spectral response (\citealp{2012arXiv1209.3114M}).
In general, due to the strong gradient in the emission measure profiles, jumps caused by shock fronts might not be as clearly visible as the temperature discontinuities, especially at large radii.}

Overall, we find that our measured profiles are very well reproduced by our deprojection method when projecting the best-fit deprojected models along the line of sight.
The reduced $\chi^2$ values in NE and SW directions are $\chi^2/{\rm d.o.f.} = 0.3$ ($\chi^2=3.4$) and $\chi^2/{\rm d.o.f.} = 0.8$ ($\chi^2=2.3$), respectively. 
The low value in the NE direction (despite of the simple deprojection model) is an indication, that the uncertainties are likely overestimated by the fit, caused by the correlations introduced by the PSF correction.

We quantify the temperature and density jumps and the resulting shock properties in the following section and give possible interpretations of the merging scenario in Sec. \ref{sec:discussion}.




%
%

\begin{figure*}[htbp]
\centering

\begin{overpic}[height=0.3\textwidth]{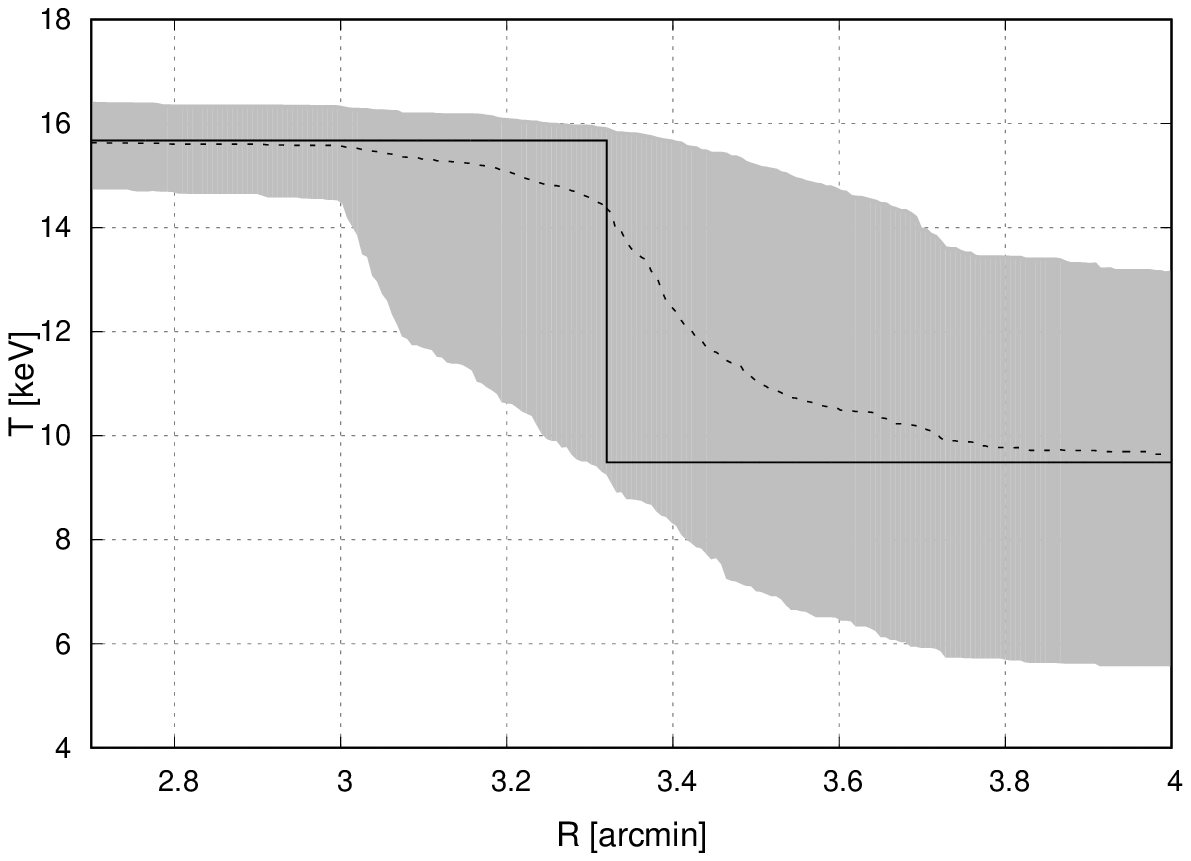}\put(87,65){\fcolorbox{black}{white}{S$_1^{\rm SW}$}}\end{overpic}
\begin{overpic}[height=0.3\textwidth]{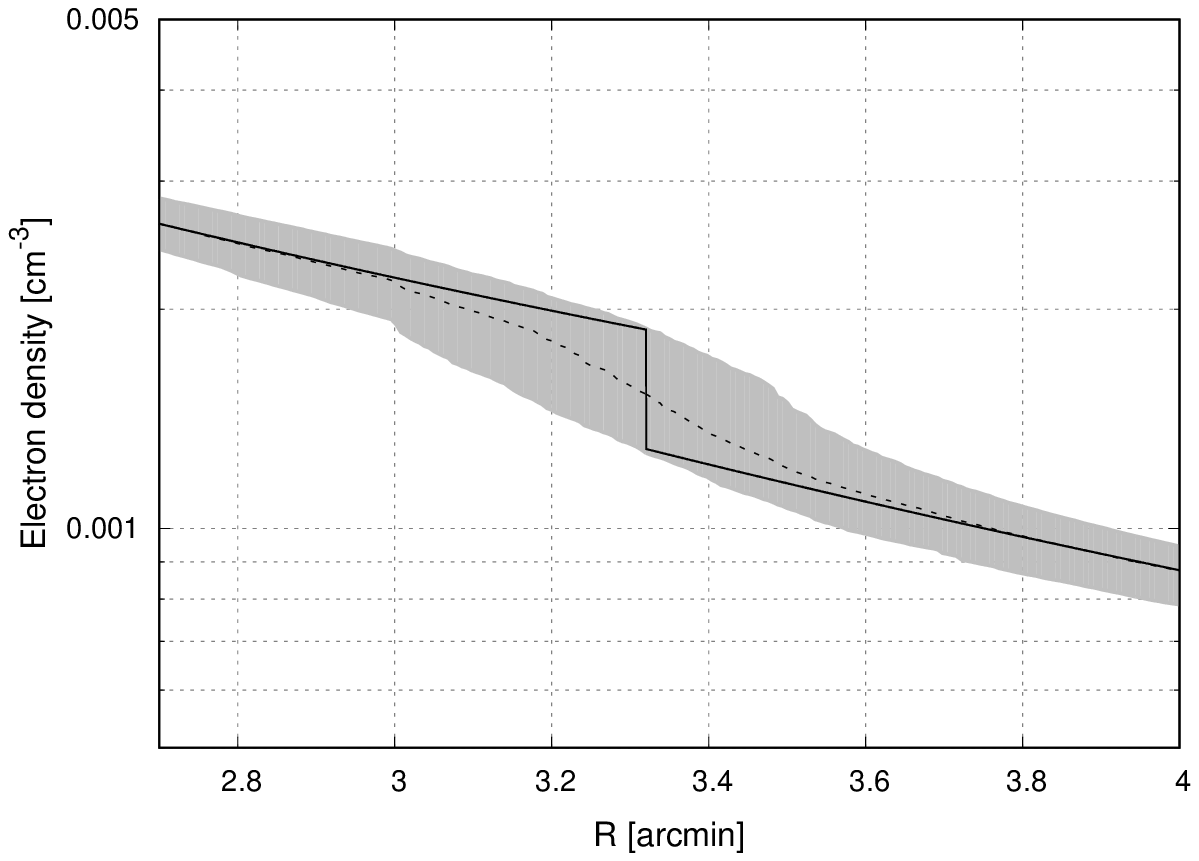}\put(87,65){\fcolorbox{black}{white}{S$_1^{\rm SW}$}}\end{overpic}
\begin{overpic}[height=0.3\textwidth]{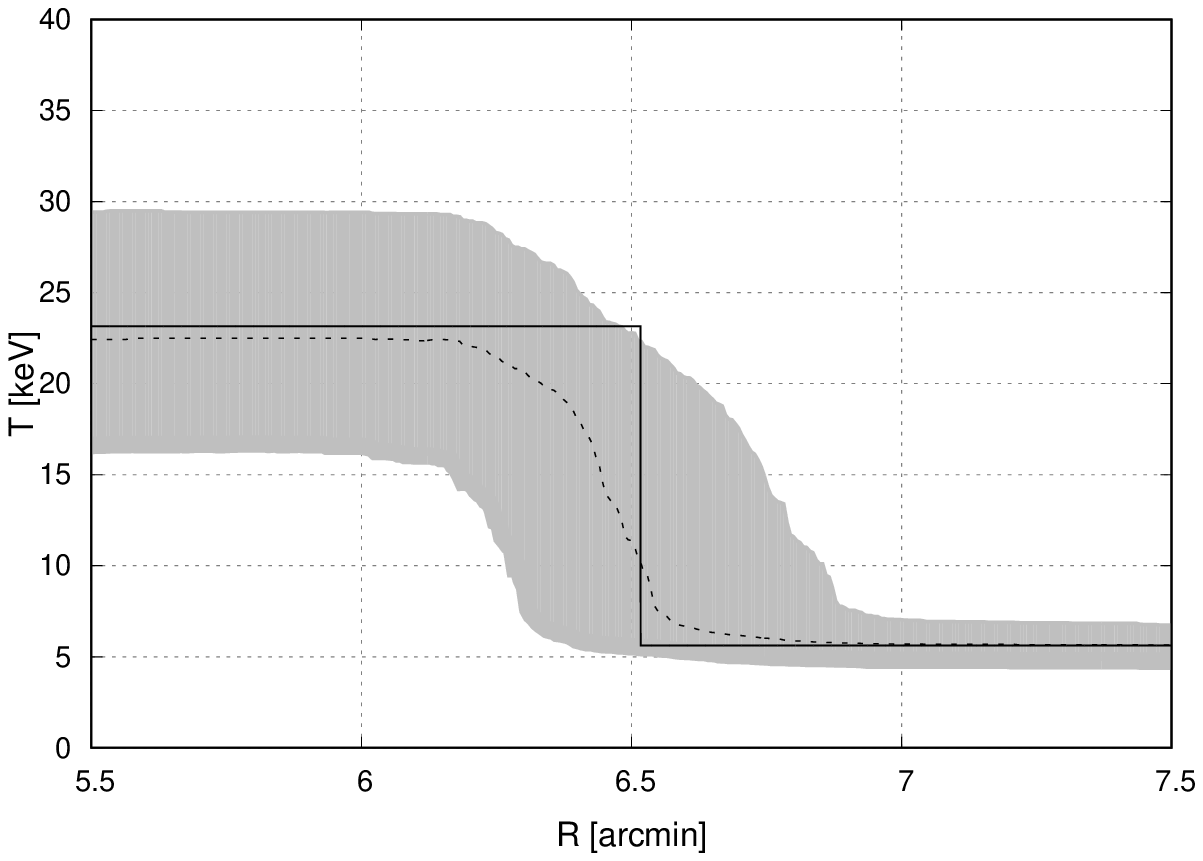}\put(87,65){\fcolorbox{black}{white}{S$_2^{\rm SW}$}}\end{overpic}
\begin{overpic}[height=0.3\textwidth]{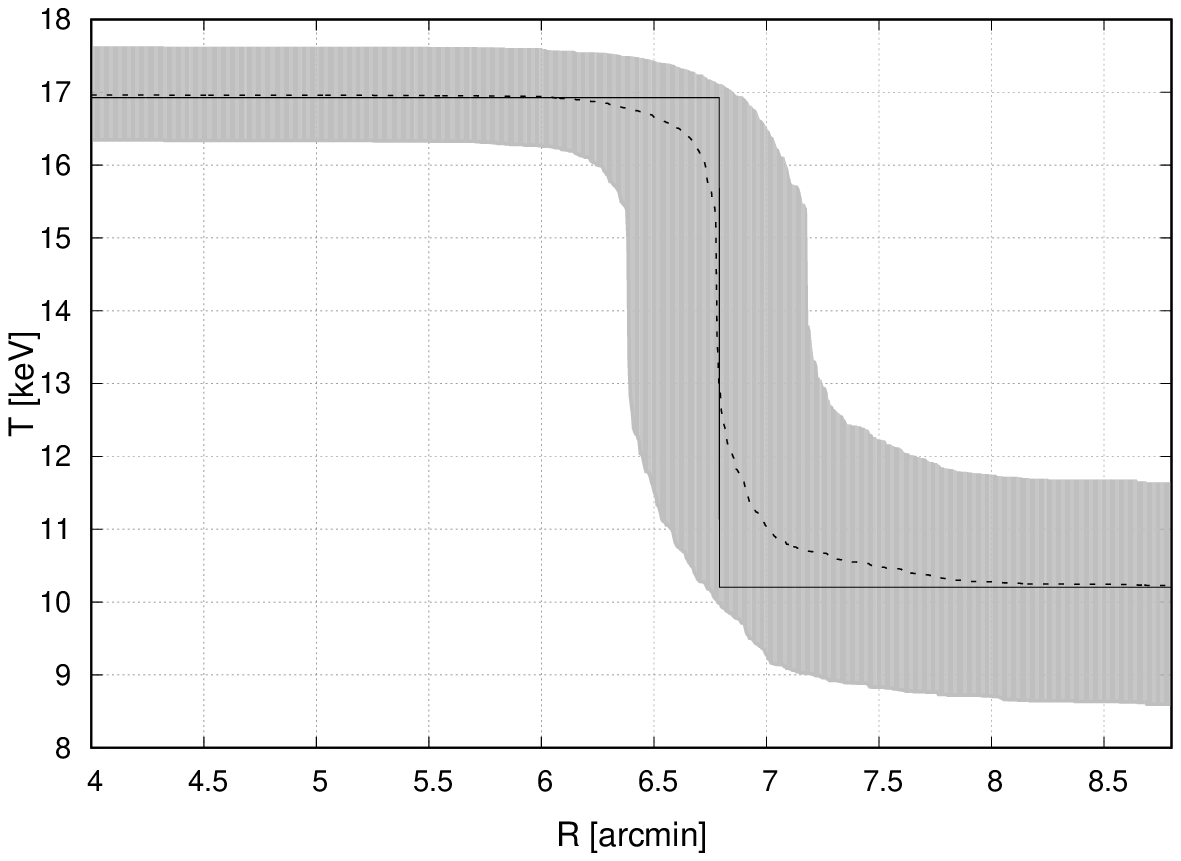}\put(87,65){\fcolorbox{black}{white}{NE}}\end{overpic}

\caption[Temperature and density jumps]{ Deprojected temperature and electron density profiles around the shock regions for the NE and SW direction. Solid lines correspond to the deprojected profiles, gray shaded areas show the 68\% confidence intervals estimated from 1000 Monte-Carlo realizations. The dashed lines show the median of all realizations. For details of the shock properties see Tab. \ref{tab:jumps}. {\it Top row:} Deprojected temperature (left) and density (right) profile around the jump position for the inner shock in SW direction.
{\it Bottom left:} Deprojected temperature profile in the shock region for the outer shock in SW direction. Note that the profile slightly exceeds the 68\% interval at the jump position because of the sharp discontinuity which is smeared out in the MC realizations. {\it Bottom right:} Deprojected temperature profile in the NE direction around the shock position. }
\label{fig:T_dens_jumps}
\end{figure*}

\subsection{Properties of the shock fronts} \label{sec:shock}
From the deprojected models, introduced in Sec.~\ref{sec:deprojection}, we estimate the jump amplitudes (i.e.\ the ratio of the respective quantity in front of and behind the shock) in the temperature and (where possible) density profiles. From these values, the Mach numbers of the shocks are estimated using the Rankine Hugoniot jump conditions. The results are given in Tab.~\ref{tab:jumps} and Fig.~\ref{fig:T_dens_jumps} shows the corresponding discontinuities in the deprojected profiles. The uncertainties are estimated from 1000 Monte-Carlo realizations of the measured profiles, for which we repeat the deprojection but reject unphysical realizations (e.g. negative temperatures). We include a $10\%$ systematic error on the spectral normalizations which comes from the uncertainty on the shape of the beta-model profile used in the ARF generation. 



\renewcommand{\arraystretch}{1.15}
\begin{table*}
\centering
\caption[A2163 shock properties]{Properties of the shock in NE and the two shocks in SW direction of A2163. $A$ is the jump amplitude, i.e.\ the ratio of the respective quantity in front of and behind the shock, $M$ is the Mach number and $v$ the shock velocity. Where the jump amplitude is smaller than one, no Mach number and velocity estimates are given.}
  \begin{tabular}{l l c | c c c | c c c }
& & & \multicolumn{3}{c |}{Density estimates}&  \multicolumn{3}{c}{Temperature estimates} \\
&$R_{\rm j}$ [arcmin]&$R_{\rm j}$ [kpc]&$A$&$M$&$v$ [$10^3$ km/s]&$A$&$M$&$v$ [$10^3$ km/s]\\\hline
S$_1^{\rm SW}$&$3.3_{-0.3}^{+0.3}$&$665_{-60}^{+55}$&$1.5_{-0.3}^{+0.3}$&$1.3_{-0.2}^{+0.2}$&$2.1_{-0.3}^{+0.4}$&$1.7_{-0.5}^{+1.3}$&$1.6_{-0.5}^{+0.9}$&$2.6_{-0.7}^{+1.5}$\\
S$_2^{\rm SW}$&$6.5_{-0.2}^{+0.3}$&$1305_{-47}^{+68}$&$0.8_{-0.1}^{+0.2}$&${-}$&${-}$&$4.1_{-1.2}^{+1.4}$&$3.2_{-0.7}^{+0.6}$&$4.0_{-0.8}^{+0.8}$\\\hline
S$^{\rm NE}$&$6.8_{-0.4}^{+0.4}$&$1360_{-78}^{+81}$&$0.8_{-0.1}^{+0.1}$&${-}$&${-}$&$1.7_{-0.2}^{+0.3}$&$1.7_{-0.2}^{+0.3}$&$2.7_{-0.4}^{+0.5}$\\
  \end{tabular}
  
%

\label{tab:jumps}
\end{table*}
\renewcommand{\arraystretch}{1.0}

The inner shock in the SW direction (labeled S$_{1}^{\rm SW}$) is located at $3.3'$ and the temperature and density profile yield consistent Mach numbers of $1.6_{-0.5}^{+0.9}$ and $1.3_{-0.2}^{+0.2}$, respectively, albeit with relatively large uncertainties. These Mach numbers correspond to shock velocities of $2100-2600$\,km/s and are typical values for X-ray detected shocks. The second shock in SW direction (S$_{2}^{\rm SW}$) as well as the shock in NE direction (S$^{\rm NE}$) are only detectable in the temperature profiles, whereas no clear discontinuities are found in the density profiles. The likely reasons for this are, on the one hand, Suzakus large PSF and, on the other hand, the steep gradient of the density profile, which makes it more difficult to detect discontinuities. 

The Mach number of the NE shock, estimated from the temperature jump, is $M=1.7_{-0.2}^{+0.3}$. The Mach number of the outer shock in SW direction is $M=3.2_{-0.7}^{+0.6}$, which is among the strongest detected X-ray shocks and comparable to the shock front detected in the Bullet cluster. However, the uncertainties are relatively large and consequently also the shock velocity has a large range of $3200-4800$\,km/s. The challenge of high collision velocities larger than ${\sim}4000$\,km/s to the $\Lambda$CDM cosmology was extensively discussed previously for the Bullet cluster (e.g. \citealp{2006MNRAS.370L..38H}, \citealp{2007MNRAS.380..911S}, \citealp{2008MNRAS.383..417A}, \citealp{2010ApJ...718...60L}). However, we do not consider such a discussion meaningful here for A2163, given the large uncertainties on the Mach numbers.

As mentioned in Sec. \ref{sec:psfcorr}, we regularized the temperature in the fitting procedure due to strong correlations caused by the PSF correction. We test the implications of this procedure for the shock properties by performing the analysis without PSF correction, i.e. without regularization. The values of this procedure are given in Tab. \ref{tab:jumps_wopsf}. The results agree well with the nominal values in Tab.~\ref{tab:jumps} within the uncertainties. The jump amplitude (and corresponding Mach number) for the strong shock S$_{2}^{\rm SW}$ in the SW direction is slightly lower, which is not surprising because the effect of the PSF smears out temperature differences in neighboring annuli. However, also these values are consistent with the regularized PSF-corrected values within the uncertainties and we thus conclude that this method is not biasing our estimated shock properties. 

\renewcommand{\arraystretch}{1.15}
\begin{table*}
\centering
\caption[A2163 shock properties]{Properties of the shock in NE and the two shocks in SW direction {\it without} PSF-correction and temperature regularization. Parameters are the same as in Tab. \ref{tab:jumps}.}
\begin{tabular}{l l c | c c c | c c c }
& & & \multicolumn{3}{c |}{Density estimates}&  \multicolumn{3}{c}{Temperature estimates} \\
&$R_{\rm j}$ [arcmin]&$R_{\rm j}$ [kpc]&$A$&$M$&$v$ [$10^3$ km/s]&$A$&$M$&$v$ [$10^3$ km/s]\\\hline
S$_1^{\rm SW}$&$3.4_{-0.3}^{+0.3}$&$680_{-53}^{+63}$&$1.2_{-0.3}^{+0.3}$&$1.2_{-0.2}^{+0.2}$&$2.1_{-0.3}^{+0.4}$&$1.3_{-0.2}^{+0.3}$&$1.3_{-0.2}^{+0.3}$&$2.3_{-0.3}^{+0.5}$\\
S$_2^{\rm SW}$&$6.6_{-0.2}^{+0.3}$&$1325_{-46}^{+60}$&$1.0_{-0.2}^{+0.3}$&${-}$&${-}$&$2.9_{-0.5}^{+0.6}$&$2.5_{-0.3}^{+0.4}$&$3.3_{-0.5}^{+0.5}$\\\hline
S$^{\rm NE}$&$6.6_{-0.4}^{+0.5}$&$1330_{-72}^{+92}$&$0.9_{-0.1}^{+0.1}$&${-}$&${-}$&$1.4_{-0.2}^{+0.3}$&$1.5_{-0.2}^{+0.3}$&$2.6_{-0.3}^{+0.4}$\\
 \end{tabular}

\label{tab:jumps_wopsf}
\end{table*}
 \renewcommand{\arraystretch}{1.0}

\section{Discussion and Conclusions}\label{sec:discussion}
\begin{figure}[htbp]
\centering
\includegraphics[width=0.49\textwidth]{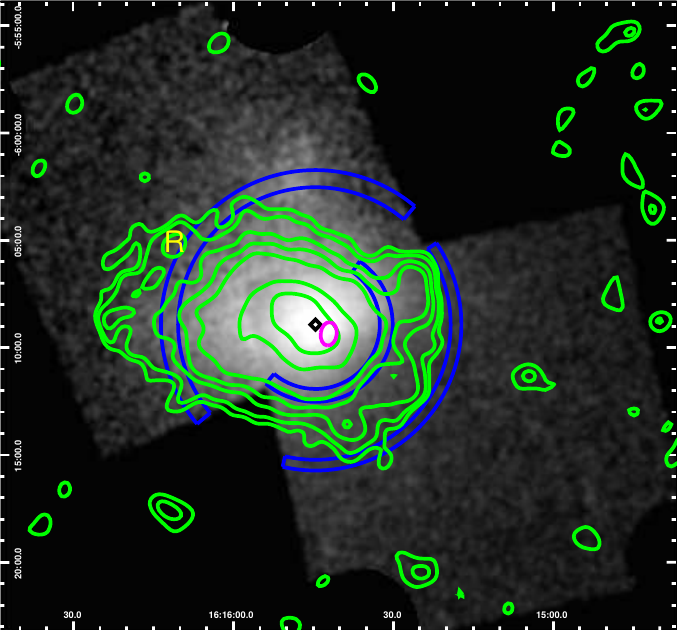}
\caption[]{Positions of the shock fronts in A2163 (blue), the X-ray emission peak (black diamond) and the approximate position of the cool core (magenta ellipse) found by  \citet{2011A&A...527A..21B}. The widths of the shock positions correspond to the uncertainties on the jump radius, estimated from the deprojection (see. Tab. \ref{tab:jumps}). Green contours show the extended radio emission with the same contour levels as in the right panel of Fig. \ref{fig:radio_img}. The yellow letter ``R'' marks the position of the radio relic as identified in \citet{2001A&A...373..106F}.
}
\label{fig:jump_positions}
\end{figure}

Fig. \ref{fig:jump_positions} shows the radio contours of A2163 as well as the estimated shock positions from this analysis. Overall, we see a tight correlation between the X-ray detected shock positions and the radio contours. A general correlation between the radio and X-ray features of A2163 was also observed previously by \citet{2001A&A...373..106F} but, so far, not associated with shock fronts.

\citet{2001A&A...373..106F} classified the radio feature in the north-east as a radio relic which spatially coincides with the position of the NE shock front. The authors also discuss the possibility that this radio feature is part of a  wide-angle-tailed (WAT) radio source with a size of  ${\sim}$1.3\,Mpc, which, however, seems unlikely as these features are usually smaller and located in the central parts of clusters. Additionally, \citet{2004A&A...423..111F} measured the spectral index map of A2163 and found that the index flattens at the peak position of the NE relic and steepens towards the edges. A flatter electron spectrum points towards a more energetic particle population. Interestingly, the spectral index map also shows an elongated region of a flatter spectral index located close to the outer SW shock front, though, this feature has not (yet) been classified as a radio relic but seems likely related to the shock front.


In the common understanding, the highly energetic electrons produced in the merger shocks are responsible for the radio emission and, in particular, can form radio relics. This was also seen previously, for example, by \citet{2011ApJ...728...82M}, \citet{2012PASJ...64...67A} and \citet{2015A&A...582A..87A} for the disturbed clusters A754, A3376 and CIZA J2242.8+5301, which all exhibit radio relics that spatially coincide with the position of X-ray detected shock fronts. 

The clumpy shape of the spectral index map of A2163, as discussed in \citet{2004A&A...423..111F}, and the flatter shape of the spectra close to the shock positions, suggests a reacceleration of the electron population at sites of recent merger activity. \citet{2004A&A...423..111F} found that in relatively undisturbed cluster regions, the index steepens with distance from the cluster center. For the interpretation of this steepening, the authors refer to \citet{2001MNRAS.320..365B}, who observed this behavior for the Coma galaxy cluster and explained it with a two-phase model, where in the first phase an energetic electron population is continuously injected by shocks, AGN activity and/or turbulences and later, in the second phase, reaccelerated e.g. by recent merging activity.


The comparison of the X-ray shock positions to the findings obtained by \citet{2011A&A...527A..21B} with XMM-Newton and Chandra shows, that the inner shock in the SW direction is located close in front of a cool spot as depicted in Fig. \ref{fig:jump_positions}. In the scenario suggested by \citet{2011A&A...527A..21B} and \citet{2008A&A...481..593M}, the merger in A2163 took place in the E-W direction (projected onto the plane of the sky) and the cold spot is likely a cool core relic of one of the merging constituents, which is moving westwards. This scenario is supported by our findings as it seems likely that the inner SW shock front is related to the cool core. Also a cold front was identified previously in a sector at the north-west boundary of the cold spot, using Chandra data (\citealp{2011A&A...527A..21B}), however, it can not be resolved with Suzaku.

\begin{figure*}[htbp]
\centering
\includegraphics[width=0.7\textwidth]{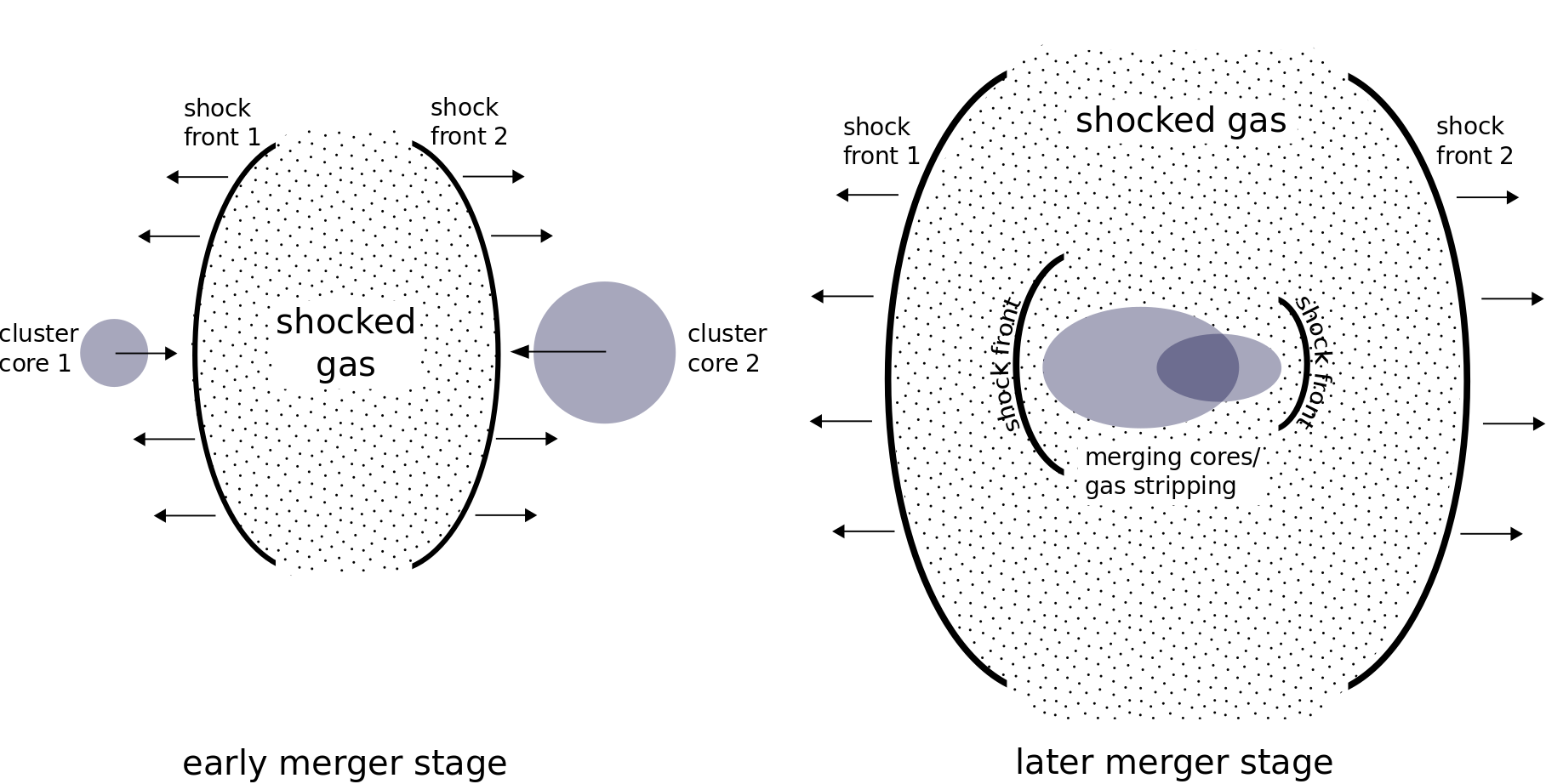}
\caption[]{Possible merger scenario adapted from \citet{2000ApJ...541..542M}. In an early stage of the merger (left), the gas in between the merging clusters gets shock heated and shock fronts develop. In a later stage (right), the cluster cores merge and can develop additional shock fronts, while the initial shock waves traveled outwards.}
\label{fig:scenario}
\end{figure*}

The presence of multiple shock fronts with somewhat different Mach numbers and shock velocities suggests a rather complex merger scenario. However, a simplified possible scenario is depicted in Fig. \ref{fig:scenario}, which was initially suggested by \citet{2000ApJ...541..542M} for the cluster A2142, but is also conceivable for A2163. Two merging constituents approach each other and develop multiple shock fronts in the ICM, which travel outwards. In an ideally symmetric head-on collision with equal masses, the velocities of these shock fronts are similar. The cores of the clusters merge and ahead of these dense (and often cool) structures, additional shock fronts can occur. This is similar to our findings for A2163, where the inner shock front in SW direction is closely situated in front of the moving cool core and two shock fronts are traveling outwards at larger radii in both directions. We find the two outer shocks at roughly equal distance from the X-ray peak. However, the Mach numbers and velocities are quite different which is likely due to deviations from an ideal head-on collision and/or unequal merging constituents. Projection effects can also play a role as it is difficult to constrain the angle between the line of sight and the merging axis, which can even lead to an underestimation of the Mach number (e.g. \citealp{2013ApJ...765...21S}, \citealp{2015ApJ...812...49H}). Additionally, the above described scenario is highly idealized and in real situations, turbulent gas motions as well as multiple minor mergers may occur. This is supported by the previously mentioned radio and optical studies, who identified several substructures in A2163. Furthermore, in the course of the merging process of the dense cores, the position of the X-ray peak can change and must not necessarily coincide with the origin of the initial shock waves, which might contribute to the fact that the two shock fronts lie at roughly the same distance from the X-ray peak despite their different velocities. 

\section{Summary}
We identified three shock fronts in A2163 using Suzaku data in two azimuthal directions and compared the X-ray findings to the radio morphology, obtained from an archival VLA observations at $20$\,cm. The NE direction exhibits one shock at at distance of $1.4$\,Mpc from the center, with a typical Mach number of X-ray detected shocks of $M=1.7_{-0.2}^{+0.3}$, estimated from the discontinuity in the temperature profile. The SW direction shows a more complicated temperature profile and we identify two shocks. One is located at $0.7$\,Mpc with $M=1.5_{-0.3}^{+0.5}$, estimated as the average from the density and temperature discontinuities; the other one lies at $1.3$\,Mpc and is one of the strongest detected shocks with $M=3.2_{-0.7}^{+0.6}$, estimated from the temperature jump. Previous studies in the radio and optical regime identified several substructures, which hints at a complicated merger scenario. The radio contours well match the positions of the X-ray detected shocks. The NE shock front coincides with the position of a radio relic, which was identified previously by \citet{2001A&A...373..106F} and \citet{2004A&A...423..111F}. Their radio spectral index map also shows a region of flatter index close to the outer SW shock front which hints at an electron reacceleration mechanism at sites of ongoing merging activity. Close to the inner shock front in the SW direction, \citet{2011A&A...527A..21B} identified a moving cool core remnant, which suggests a connection between the two. The merging scenario can qualitatively be explained by two merging constituents, which passed each other and developed shock fronts in an early stage of the merger. These shock fronts traveled outwards, while the dense cluster cores merged and formed additional shock and cold fronts ahead of them. The different Mach numbers and shock velocities, as well as the substructures identified in the radio and optical regime, suggest that the real merger situation is more complex and likely occurred off-axis with unequal merging constituents.


\begin{acknowledgements}
ST acknowledges support from the Bonn-Cologne Graduate School of Physics and Astronomy and the Argelander-Institut f\"ur Astronomie. THR acknowledges support by the German Aerospace Agency (DLR) with funds from the Ministry of Economy and Technology (BMWi) through grant 50 OR 1514. MWS acknowledges support for this work from Transregio Programme TR33 of the German Research Foundation (Deutsche Forschungsgemeinschaft). This work was supported in part by JSPS KAKENHI Grant Number 16K05295 (NO).
\end{acknowledgements}

\bibliographystyle{aa} 
\bibliography{bibtex}
\begin{appendix}
 \section{}

 \begin{table}[htbp]
\centering
\caption[Fit results in the NE direction]{Fit results in the NE direction.}
\begin{tabular}{c | c | c | c}
 Annulus [arcmin] & $T$ [keV] &   $Z$ [$Z_\odot$] &   norm$^1$ \\ \hline\hline 

$ 0-1 $&$ 14.13 \pm 1.65 $&\multirow{ 1}{*}{$ 0.68 \pm 0.17 $}&$ 76.53 \pm 0.94 $\\\hline
$ 1-2 $&$ 17.01 \pm 1.96 $&\multirow{ 2}{*}{$ 0.33 \pm 0.11 $}&$ 151.91 \pm 0.98 $\\\cline{1-2} \cline{4-4}
$ 2-3 $&$ 16.03 \pm 2.89 $&$    $&$ 75.36 \pm 0.50 $\\\hline
$ 3-4 $&$ 17.87 \pm 5.12 $&\multirow{ 3}{*}{$ 0.60 \pm 0.15 $}&$ 43.37 \pm 0.40 $\\\cline{1-2} \cline{4-4}
$ 4-5 $&$ 14.88 \pm 2.46 $&$    $&$ 35.70 \pm 0.45 $\\\cline{1-2} \cline{4-4}
$ 5-6 $&$ 19.16 \pm 5.23 $&$    $&$ 23.08 \pm 0.37 $\\\hline
$ 6-7 $&$ 10.38 \pm 2.13 $&\multirow{ 3}{*}{$ 0.20 \pm 0.16 $}&$ 20.34 \pm 0.36 $\\\cline{1-2} \cline{4-4}
$ 7-8 $&$ 10.71 \pm 5.00 $&$    $&$ 15.16 \pm 0.39 $\\\cline{1-2} \cline{4-4}
$ 8-9 $&$ 13.33 \pm 6.74 $&$    $&$ 9.81 \pm 0.35 $\\\cline{1-2} \cline{4-4}
$ 9-10 $&$ 9.76 \pm 4.24 $&$    $&$ 6.03 \pm 0.32 $\\\hline

\end{tabular}
\label{tab:fitresults_NE}
\begin{minipage}{0.49\textwidth}
\vspace{0.2cm}
{\tiny $^1$ normalization in the full annulus: ${\rm norm}=\frac{10^{-18}}{4\pi[D_A(1+z)]^2}\int n_{\rm e}n_{\rm H}{\rm d}V\,{\rm cm}^{-5}$ with $D_A$ being the angular diameter distance to the source.}
\end{minipage}
\end{table}

\begin{table}[htbp]
\centering
\caption[Fit results in the SW direction]{Fit results in the SW direction.}
\begin{tabular}{c | c | c | c}
Annulus [arcmin] & $T$ [keV] &   $Z$ [$Z_\odot$] &   norm$^1$ \\\hline\hline

$ 0-1 $&$ 14.21 \pm 1.25 $&\multirow{ 1}{*}{$ 0.57 \pm 0.11 $}&$ 84.26 \pm 0.48 $\\ \hline
$ 1-2 $&$ 15.01 \pm 1.46 $&\multirow{ 2}{*}{$ 0.31 \pm 0.07 $}&$ 149.22 \pm 0.52 $\\ \cline{1-2} \cline{4-4}
$ 2-3 $&$ 15.53 \pm 1.32 $&$    $&$ 64.80 \pm 0.30 $\\ \hline
$ 3-4 $&$ 12.09 \pm 1.42 $&\multirow{ 3}{*}{$ 0.12 \pm 0.12 $}&$ 22.91 \pm 0.19 $\\ \cline{1-2} \cline{4-4}
$ 4-5 $&$ 13.59 \pm 2.70 $&$    $&$ 11.84 \pm 0.17 $\\ \cline{1-2} \cline{4-4}
$ 5-6 $&$ 16.84 \pm 3.94 $&$    $&$ 7.98 \pm 0.17 $\\ \hline
$ 6-7 $&$ 8.08 \pm 1.21 $&\multirow{ 3}{*}{$ 0.31 \pm 0.12 $}&$ 6.03 \pm 0.19 $\\ \cline{1-2} \cline{4-4}
$ 7-8 $&$ 7.41 \pm 1.55 $&$    $&$ 4.36 \pm 0.20 $\\ \cline{1-2} \cline{4-4}
$ 8-9 $&$ 4.73 \pm 1.12 $&$    $&$ 2.87 \pm 0.20 $\\ \hline

\end{tabular}
\label{tab:fitresults_SW}
\begin{minipage}{0.49\textwidth}
\vspace{0.2cm}
{\tiny $^1$ same definition as in Tab. \ref{tab:fitresults_NE}.
}
\end{minipage}
\end{table}

\end{appendix}

\end{document}